**Title**: Applying computational protein design to therapeutic antibody discovery - current state and perspectives

**Authors**: Weronika Bielska[1,2†], Igor Jaszczyszyn[1,3†], Pawel Dudzic[1], Bartosz Janusz[1], Dawid Chomicz[1], Sonia Wrobel[1], Victor Greiff[4,5], Ryan Feehan[6], Jared Adolf-Bryfogle[6], Konrad Krawczyk [1,*]

* correspondence: konrad@naturalantibody.com

1 NaturalAntibody
2 Medical University of Lodz, Lodz, Poland
3 Medical University of Warsaw, Warsaw, Poland
4 Department of Immunology, University of Oslo, Oslo, Norway
5 Imprint Labs, LLC. New York, NY, USA
6 Janssen Pharmaceuticals

† - These authors contributed equally to this work and share first authorship

**Abstract**: Machine learning applications in protein sciences have ushered in a new era for designing molecules in silico. Antibodies, which currently form the largest group of biologics in clinical use, stand to benefit greatly from this shift. Despite the proliferation of these protein design tools, their direct application to antibodies is often limited by the unique structural biology of these molecules. Here, we review the current computational methods for antibody design, highlighting their role in advancing computational drug discovery.

Introduction

Antibodies represent the largest class of biotherapeutics [1], demonstrating significant versatility and efficacy in treating a wide array of diseases, including cancer, autoimmune disorders, and infectious diseases. These Y-shaped proteins, also known as immunoglobulins, possess the unique ability to specifically bind to antigens, thereby marking them for destruction or neutralization by the immune system. The specificity and high binding affinity of antibodies make them invaluable tools in both therapeutic and diagnostic applications.

Traditionally, the discovery and development of therapeutic antibodies have relied on two experimental paradigms: immunization and display technologies [2]. The immunization approach involves the administration of an antigen into a host animal, such as mice or rabbits, to elicit an immune response. This process leads to the generation of polyclonal antibodies, from which monoclonal antibodies can be derived through hybridoma technology. Köhler and Milstein's pioneering work in the 1970s on hybridoma technology revolutionized antibody production by enabling the creation of monoclonal antibodies with defined specificity and uniform characteristics [3].

In contrast, display technologies, such as phage display, yeast display, ribosome display and mammalian display [4], have emerged as powerful tools for antibody discovery without the need for immunization. These methods involve the presentation of vast libraries of antibody variants on the surface of bacteriophages, yeast cells, ribosomes or mammalian cells, respectively. Through iterative rounds of selection and amplification, antibodies with high affinity and specificity for a target antigen can be isolated. Phage display, in particular, has been instrumental in the discovery of several clinically approved antibodies, with Smith's 1985 innovation marking a significant milestone in this field [5]. However, this technology has inherent limitations. Because bacterial folding machinery does not readily support the production of full-length antibodies, phage display is often limited to smaller constructs such as single-chain variable fragments (scFvs). Moreover, controlling post-translational modifications in microbial expression systems is challenging, an issue resolved by using mammalian display systems [6]. While display technologies help circumvent the need for direct immunization (and in many cases can be used in tandem for affinity or specificity improvement), it is worth noting that in vivo approaches—such as immunization—also allow receptor editing processes that can reduce the likelihood of autoreactivity.

Together, these traditional methods have laid a robust foundation for antibody discovery. However, they also present limitations, such as time-consuming processes and dependence on the host immune response or large library sizes. To address these challenges, computational antibody design has emerged as a promising complementary approach. It leverages advances in computational biology, structural bioinformatics, and artificial intelligence to expedite and enhance antibody development [7].

At its core, Computational Antibody Design is a sub-problem of the more generalistic Computational Protein Design (CPD), that aims to engineer novel proteins with desired functions and properties. CPD involves the prediction and optimization of protein structures and sequences to achieve specific functional outcomes. Key early methods in CPD include de novo design, homology modeling, and molecular dynamics simulations. De novo design involves creating novel protein structures from scratch, guided by principles of protein folding and stability [8]. Homology modeling, on the other hand, predicts structures based on the alignment with known homologs, facilitating the design of proteins with altered functions while maintaining structural integrity [9]. Molecular dynamics simulations provide insights into the dynamic behavior of proteins, allowing for the refinement of models and prediction of their stability and interactions under physiological conditions [10]. Because of reliance on structural information, such early design methods mostly used structural fragments, energy functions and statistical potentials to design new structures and sequences [11–14].

Recent advancements in machine learning-based structure [15,16] and sequence prediction [16] have given a major boost to CPD. Thanks to advancements in structure prediction spearheaded by AlphaFold2 [15], three-dimensional structures have become much more accessible [17]. Merging learnings from machine learning on natural language with protein sequences resulted in large language models such as ESM that can accurately model the distribution of natural sequences, to generate new ones. Specifically, the shift towards a 'generative' paradigm in protein and thus antibody design is the most prominent. As much as earlier methods relied on assembling fragments of known proteins, novel tools such as RFDiffusion [18], ProteinMPNN [19] or ESM-IF [20] can generate novel structures/sequences that were not observed in nature, but guided by it. Such methods are increasingly being applied to antibodies [21–23], and we provide a review and a perspective on the field here.

**Computational Protein Design Primer**

Computational protein design is crucial for developing novel biotechnological applications such as new therapeutics or industrial enzymes [24,25]. Computational protein design predominantly uses methods that leverage physicochemical calculations or machine learning to perform tasks ranging from single point mutations with increased activity to de novo design of highly thermostable proteins. Computational protein design strategies can be loosely categorized into three overlapping groups, template based protein design given structure, sequence optimization given sequence or structure and finally de novo design.

Template-based protein design relies on using existing protein structures as starting points to guide the design process - for both sequence and backbone redesign. Since protein structure determines function, this approach is particularly effective for designing proteins with new functions or enhancing existing ones. An instrumental piece of software in this sphere is Rosetta [26]. Rosetta is a software suite for molecular modeling and design with a wide range of applications that are centered around the use of protein structure and a scoring function, made up of empirical and physicochemical terms. The simplest form of computational design with Rosetta [27] is optimizing a protein's function by identifying mutations that improve its energy score.

Historically, template-based design has been limited to proteins with solved structures of closely related homologs. Recent developments in methods using ML have significantly expanded the number of use cases for computational protein design that leverage protein structures as input. Previously, starting points for designs were limited to proteins with experimentally solved structures in the PDB [28], or close homologs that could be modeled from those structures. The ability to make high-quality, computationally generated protein structures increases the number of starting structures from ~200,000 available proteins in the PDB to 200 million known protein structures in the AlphaFold database [17]. Moreover, the predicted structures of designed sequences can be used to filter out poor designs using the predicted structure's confidence metrics or by aligning the predicted structure to the designed structure. It should be noted that co-folding the interaction between two proteins to use as starting templates using tools such as Alphafold-Multimer [29] is still a very difficult challenge and even more difficult for antibody-antigen interactions.

Such large numbers of predicted structures improve the power of sequence optimization algorithms. Here, given a structural template, one is tasked with developing a sequence that would 'fit' into it (i.e. maximize the probability of sequence given structure). Current sequence optimization strategies typically take the form of inverse folding, where algorithms such as ESM-IF [20] or ProteinMPNN [19] trained on millions of predicted structures are tasked with returning the original sequences. Both ESM-IF and ProteinMPNN use a graph architecture to turn information about residues in the local neighborhood of a specific position into features for that position [19,20]. Using a message-passing neural network (MPNN) in an iterative fashion allows features at each residue position to encode information about the microenvironment of the neighboring residues. A decoder uses the structure-based embedding to generate a protein sequence that is likely to successfully fold into the input protein structure. A common evaluation for protein design tools is to calculate the sequence recovery rate, which is the percent of generated residues that match the native amino acid at that position. ESM-IF achieves 51% sequence recovery [20], while ProteinMPNN achieved 53% sequence recovery rate [19]. That is a significant improvement over Rosetta's 33% sequence recovery rate for the same proteins. Moreover, experimental validation was used to show ProteinMPNN can successfully rescue previous failed designs, increase stability, increase solubility, and even redesign membrane proteins to be available in solution [30].

In contrast to template-based and sequence-optimization methods that require the existence of a basis structure or starting sequence, de novo protein design involves creating entirely new folds from scratch. Traditional approaches, grounded in physics-based modeling, use atomistic representations and energy functions to optimize sequences for a defined protein backbone [31]. These methods rely on iterative cycles of structure generation and sequence optimization, as exemplified in early successes like the first de novo protein design of Top7 [32]. Advancements in methods using diffusion models have further expanded the potential for computational protein design by generating protein backbones that are different (but inspired by) those found in nature. For instance, RFDiffusion [18] learned to sample the large conformational landscape of protein structure by training to recover solved protein structures corrupted with noise. During inference, unconstrained predictions transform random noise into proteins that can have little overall structural similarity to any known protein structure. Additionally, RFDiffusion can be constrained with a given active site, motif, or binding partner, which enabled successful computational designs of de novo protein binders with higher rates of success than previous methods. These tools emphasize modularity, tunability, and precision; facilitating the design of proteins with programmable behaviors for applications in catalysis, molecular recognition, and synthetic biology [33,34].

Computational protein design is currently undergoing an exciting transition from predominantly energy-based methods to those using machine learning. The recent developments and success of the field have been emphasized by the Nobel Prize in Chemistry 2024 awarded for computational protein design and structure prediction to David Baker, John Jumper, and Demis Hassabis [35]. A large area of interest for protein design is the development and optimization of protein therapeutics. While many protein families can act as drugs, such as enzymes and cytokines, antibodies are the most widely used class of biologics owing to their quasi-programmable nature [36]. The convergence of generic protein design methods with therapeutic antibody discovery presents a promising avenue for translating advancements in protein design into therapeutic applications.

**Specifics of antibody structure and function for protein design**

Antibodies are proteins of the immune system that have evolved in jawed vertebrates to recognize foreign pathogens and facilitate their expulsion from the organism. They are the actuators of the adaptive immunity, as opposed to innate immunity mediated mostly by T-cell receptors. Though they are versatile binders, they are much more structurally constrained than general proteins (Figure 1A,B), which introduces nuances in the way that protein design methods addressing them need to be adjusted.

Each organism has millions of distinct antibodies that collectively represent molecular diversity that should be capable of weakly binding a non-self antigen to start an immune response [37]. The ability of antibodies to recognize virtually limitless amounts of antigens is the key to their success and of interest for protein design. Nature evolved antibodies to have their binding site composed of six complementarity determining regions, housed in a largely invariant framework (Figure 1C). Minute changes between CDRs can radically alter the binding affinity and specificity [38,39]. For this reason, whilst general de novo protein design might focus on building the entire scaffold that could interact with a binding partner, in case of antibodies, roughly 80% of the sequence should be known a priori because of the relative invariability of the framework. Much of the focus for antibody redesign is devoted to the CDR-H3, since it is the most variable and in many cases, confers most of the binding affinity and specificity.

Though most antibody design is focused on the CDR regions, it is known that the framework also has some influence on the binding ability [22,23,40]. For humanization, one needs to replace the murine framework with a human one (Figure 2A), whilst maintaining high-affinity binding [41]. This becomes a reverse design problem to focusing on CDRs alone, as one seeks to find a human framework that would be most structurally suitable to house the novel CDRs. Because of the relative invariability of the framework, one can often start from an existing binder and re-design the CDRs in one-shot fashion (Figure 2B). Arguably, the more difficult task is de novo design, when given a target antigen and epitope, one needs to create a whole new antibody that binds specifically to this epitope (Figure 2C).

Though a large unsolved part of antibody design is developing a binder, much of the preclinical work in antibody discovery is spent not on finding the right binder but on tuning the overall properties of the antibody to be more favorable as a therapeutic. These properties are commonly referred to as *developability properties* [42–44]. This is an umbrella term encompassing multiple biophysical properties that ensure that an antibody can be economically produced in necessary quantities, can be stored for a defined period of time, and has a non-risky profile from pk/pd, specificity, and toxicity point of view, before eventually moving to clinical trials. Here, optimization takes multiple forms, with both CDRs and frameworks becoming the targets for re-design. Nevertheless, in the most widely used meaning of the term antibody design we mean developing or redeveloping a binder towards a specific antigen first, modulating developability properties second.

To introduce a level of ontology into the antibody design field, we divided the methods into a number of categories, depicted in Figure 3 with details in Table 1. The methods vary in the way they accept input (structure or sequence) and what output they produce (ready-to use antibody or just a scaffold). This is not to say that some methods are incomplete; rather they have different applications within antibody design, as laid out in the following sections.

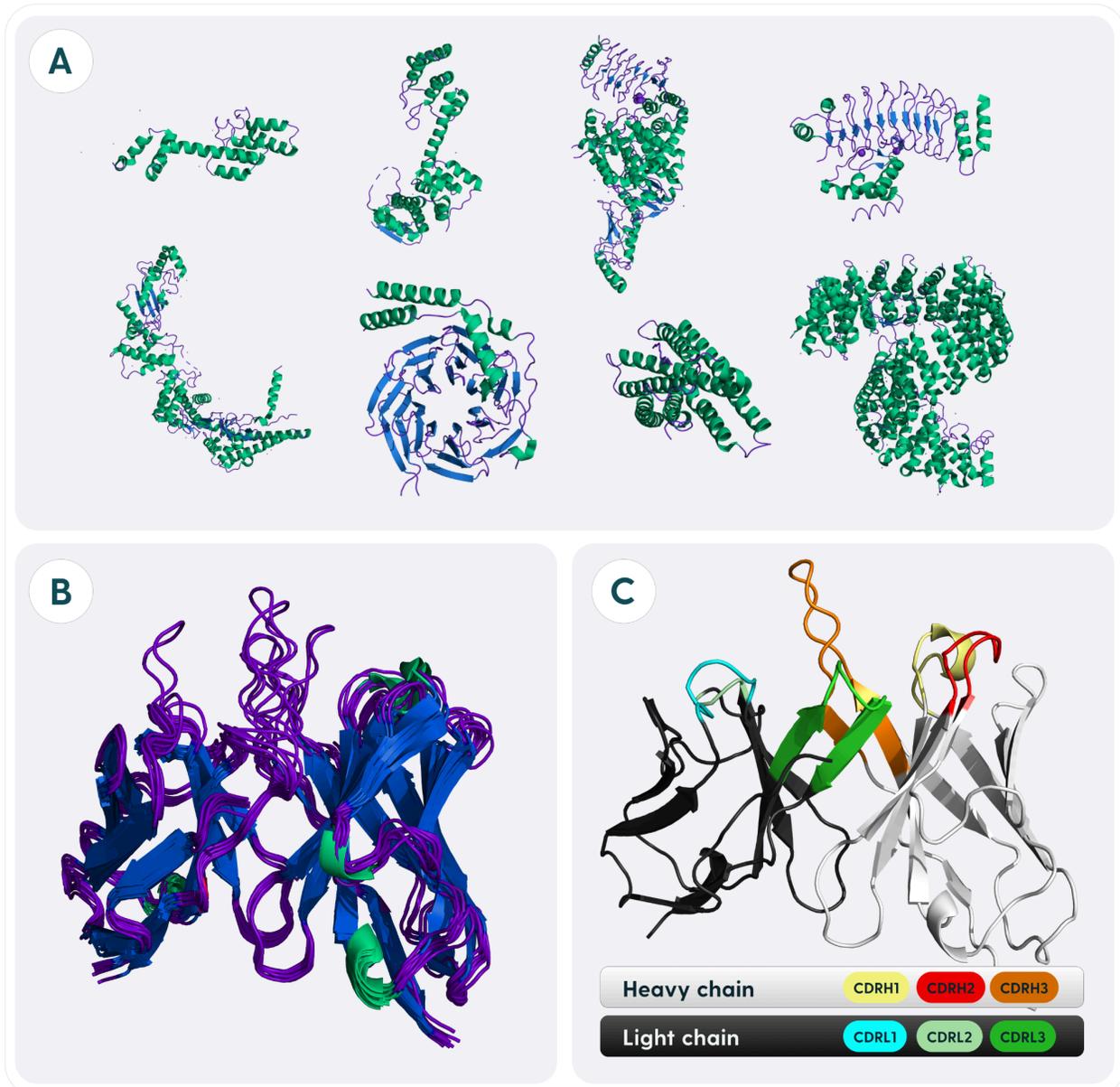

**Figure 1. Specifics of antibody structure relating to its designability as opposed to other proteins. A. Structural heterogeneity of proteins.** Proteins in general adopt a variety of conformations. Relations between folds can be drawn on an evolutionary level from sequences alone. **B. Structural homogeneity of antibodies.** Antibodies have a very conserved fold with a framework housing a diverse binding site. The differences between any two antibodies cannot be explained evolutionarily as it is the case with most proteins. **C. Regions of antibodies responsible for antigen-recognition.** Antibodies are divided into a heavy chain, light chain. Each chain is composed of three Complementarity Determining Regions (CDRs) and four Framework Regions (FR).

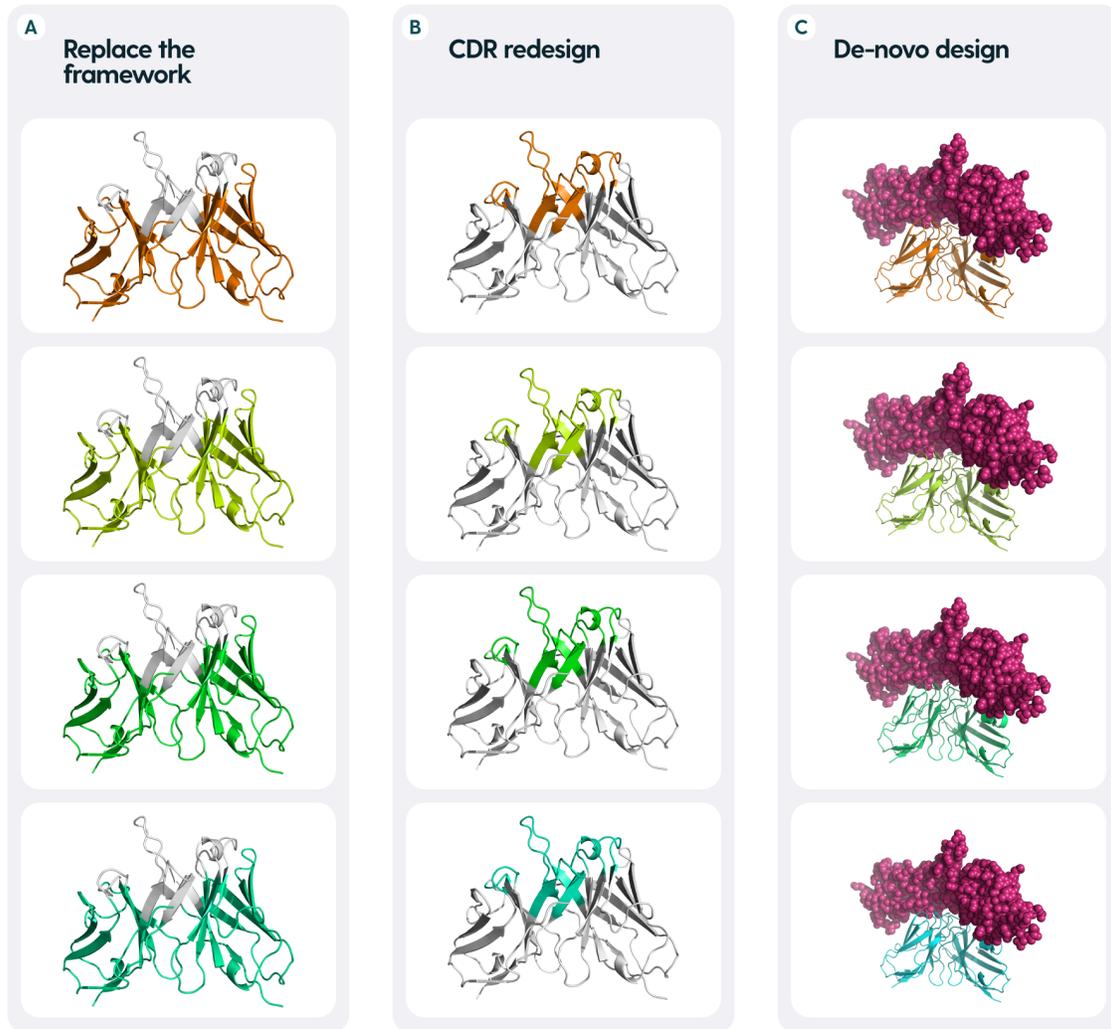

**Figure 2. Common tasks in antibody design. A.** Re-designing frameworks aim to maintain the binding of the CDRs, whilst optimizing for properties such as stability or smaller immunogenicity. **B.** Re-designing CDRs is chiefly aimed at modulating the binding abilities - specificity and affinity, usually starting from a known binder **C.** De-novo design aims to create a novel antibody molecule from the ground up, given an antigen and/or an epitope site to be targeted.

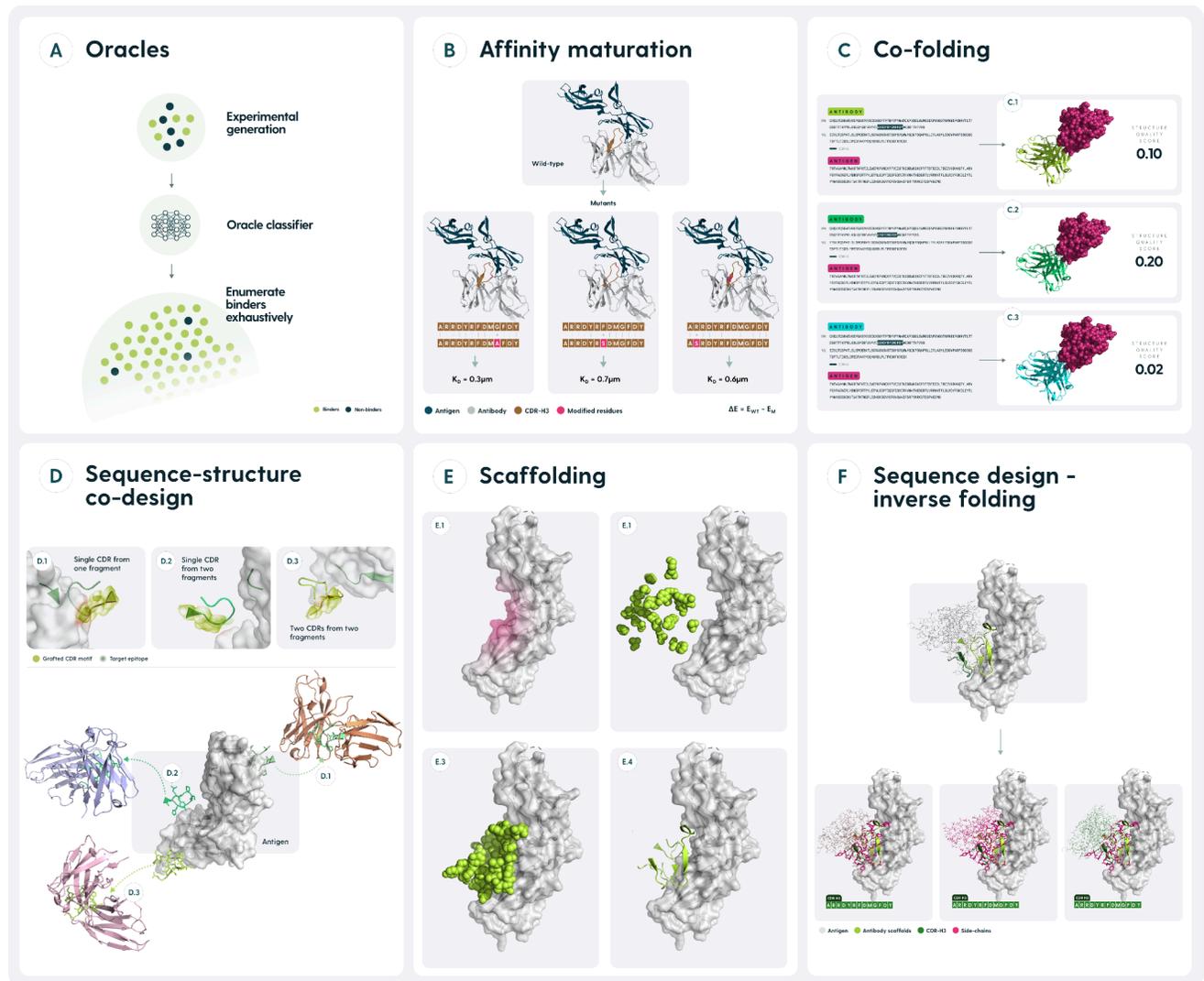

**Figure 3. Current approaches to design antibody binders computationally. A.** Binders and non-binders against a target are generated experimentally. Subsequent prediction of the binder/non-binder class allows for much more comprehensive sampling of the entire design space. **B.** Affinity maturation approaches predict the free energy changes of mutations, chiefly to the CDRs so as to obtain a larger set of binding antibodies. **C.** Current co-folding structure predictors provide confidence scores for the model of the entire complex which can be used as a proxy to gauge whether the two molecules would interact. **D.** Sequence & structure can be co-designed from pre-existing elements, such as CDR fragments or entire canonical CDRs. **E.** Backbone structure of an antibody binding a target epitope can be obtained, typically by diffusion approaches. Such predictions require a follow-up in the form of inverse folding. **F.** Given a structure of an antibody, predict a sequence that could fold into it. Applicable as a follow up to scaffolding approaches or to obtain larger sets of potential binders that have the same structure.

**Maintaining binding function of the parent antibody - efficient exploration of the binder space without knowledge of the antigen in zero-shot fashion.**

If we already have an antibody that binds to a target, we can employ it to generate more binders of antibodies. Different methods exist depending on the availability of associated experimental data - i.e. structure-enabled or not. For purely sequence-based tasks, we can use the starting antibody sequence to explore evolutionarily plausible mutations. This approach is cognate to general protein language models, where one learns to explore the fitness landscape of sequence-function relationships [45]. For instance, models such as ProGEN or ESM3, can be used for generating novel sequences that maintain certain functions [46,47] (e.g. fluorescence). Maintaining binding in antibodies is not a trivial task as introducing single mutations or combining favorable mutations is not guaranteed to maintain binding [12]. Therefore, exploring the space of 'favorable' mutations is desirable for binder development.

The approach was pioneered by Hie et al. [23] where the authors used the ESM-1b language model and the ESM-1v ensemble of five models (six models in total) to guide the evolution of seven antibodies targeting viral antigens such as SARS-CoV-2, Ebola, and Influenza A. Mutations were introduced based on the evolutionary likelihood of single-residue substitutions in the antibody variable regions (VH and VL), with substitutions that had higher evolutionary likelihood than the wild-type selected. A consensus of the six models was used to identify the most plausible substitutions. In the first round of evolution, variants with single-residue substitutions were experimentally tested for improved binding, and in the second round, combinations of beneficial substitutions were introduced to further enhance antibody affinity. Interestingly, most of the mutations recommended by the models occurred in the framework regions rather than the complementarity-determining regions (CDRs), with around half of the affinity-enhancing mutations located in these typically less mutated regions.

The aforementioned model had no notion of three dimensional structure - this was addressed in ProseLM [48], where structural adapter layers were introduced to include three dimensional information. ProseLM, builds upon the Progen family of models, incorporating structural information to improve the design of therapeutic antibodies. This structural information is integrated through structural adapter layers added after the language model layers, encoding backbone details and associated functional annotations. Models with more parameters show significant improvements in perplexity, with further gains observed when incorporating additional context information such as ligands. An antibody-specific version of ProseLM was trained exclusively on the SABDAB [49] dataset, achieving superior sequence recovery performance compared to larger models. The model was used to propose mutations for therapeutic antibodies Nivolumab and Secukinumab, targeting both the CDRs and framework regions, with designs based on structures from the PDB. Experimental results revealed that redesigning frameworks led to a much higher success rate in maintaining binding (92%), while redesigning CDRs resulted in a lower success rate (25% for Nivolumab).

A notable study that performed large-scale validation is by Shanehsazzadeh et al. [50]. The authors employed their AI model to generate CDR-h3s and all-CDR variants of trastuzumab. Care was taken to remove all close-homologs of trastuzumab.  The model focused on generating heavy chain complementarity-determining regions (HCDRs) in a zero-shot fashion - without prior exposure of the model to the target antigen. The study generated a library of about 400,000 HCDR variants and validated binders using high-throughput surface plasmon resonance (SPR) experiments. The results identified 421 diverse binders, with 71 showing low nanomolar affinity to HER2, and some antibodies performing on par or better than the therapeutic antibody trastuzumab. The top-performing generative models significantly outperformed biological baselines such as OAS and SAbDab databases, achieving a 10.6% binding rate for HCDR3 designs and a 1.8% rate for full HCDR123 designs.

The tool of choice for the few shot design is a language model that was trained using autoregressive or masking procedure. Although there is already a fair number of language models and their antibody-specific varieties [51–53], there were not multiple zero-shot/few-shot exercises like the ones above. Though such approaches appear to be well suited to explore the space around a specific binder, they do not offer a way to radically deviate from it. Therefore it is also desirable to have a more precise set of binders against a given target, which can be achieved by a combination of data-generation and supervised learning.

**Oracles - Large-scale binder generation and subsequent machine learning model training.**

One of the main tasks of an antibody design exercise is to develop a binder towards an antigen. Machine learning methods are notoriously data-intensive so an approach that has been explored by some groups was to generate prediction-first machine learning datasets of binders and non-binders and train the models on these (Figure 3A).

Recent advancements in computational antibody design have leveraged high-throughput experimental data to train machine learning (ML) models, achieving remarkable success in predicting antigen specificity and binding affinity. Mason et al. [38] pioneered this approach by deep-sequencing libraries of trastuzumab variants and training a convolutional neural network (CNN) to predict HER2 specificity, achieving an area under the ROC curve (AUC) of 0.91. Similarly, Lim et al. (2022) [54] generated datasets for antibodies targeting CTLA-4 and PD-1 by sorting yeast-displayed libraries, with their CNN achieving AUC values of 0.90 and 0.94, respectively. Building on this foundation, Chinery et al. (2024) [39] expanded the scope, creating a dataset of over 524,000 trastuzumab variants classified by binding affinity to HER2 and benchmarking multiple ML models, including CNNs, Fast Library for Automated Machine Learning (FLAML), and Equivariant Graph Neural Networks (EGNN). Notably, the CNN excelled in low-data scenarios, while FLAML performed better on larger datasets. This work also integrated computational methods like AbLang and ProteinMPNN to enrich high-affinity variants, underscoring the potential of ML in optimizing antibody libraries with efficiencies comparable to traditional experimental methods.

The above-mentioned models were trained on datasets generated for this purpose. However, it is believed that fine-tuning models offers much improvement [55]. Here, fine-tuning is understood as taking a feature-representation model, such as a language model, trained on many unrelated antibody/protein sequences, and focusing it on a library of antibody-specific ones.

For instance, Engelhart et al. (2022) [56] generated the AlphASeq SARS-CoV-2 dataset of 104,972 antibody sequences with quantitative binding data, enabling Deutschmann et al. (2024) [57] to fine-tune and benchmark domain-agnostic and domain-specific models. The ESM2 model outperformed AbLang in predicting binding affinities, demonstrating the power of generalist models when trained on large datasets. Similarly, Krause et al. (2023) [58] fine-tuned ProGen with 60 CD40-targeting antibodies to bias sequence generation toward improved affinity. Barton et al. (2024) [59] introduced FAbCon, a generative antibody-specific language model fine-tuned on datasets like AlphASeq and others, achieving state-of-the-art predictive performance (e.g., AUROC of 0.815 for SARS-CoV-2 binding) and generating low-immunogenicity antibodies validated through computational developability assessments. Finally, AlphaBind [60] utilized pre-training on 7.5 million affinity measurements and fine-tuning on experimental data, incorporating sequence embeddings from ESM-2nv to optimize antibodies for binding affinity and developability. These studies highlight how fine-tuning enhances ML models' ability to predict and generate optimized antibodies for diverse therapeutic targets.

Altogether the DMS-based methods demonstrate that it is possible to train machine learning models if enough data is available (according to Lim et al. in the order of hundreds of binders/non binders is enough). Such approaches require generating a large experimental dataset to then train a neural network. The overhead is justified by the subsequent ability to computationally scan a much larger space of binders, in search of antibodies with better developability or binding properties. Such approaches are paradoxically antigen-specific but do not require antigen at prediction time. The methods chiefly learn the distribution of the antibody-side, or just the CDR-H3 that recognizes the antigen. Therefore each is very constrained to the DMS antigen, lacking generalizability. The overarching task of antibody design remains to be able to generalize to any kind of antigen at the start. Such design is typically approached by structure-based and de novo methods.

**Affinity maturation/structure optimization.**

Given the importance of enhancing antibody-antigen binding affinity, computational methods for affinity maturation have evolved significantly over the years. Early approaches, such as those by Lippow et al., utilized physics-based energy functions like CHARMM to systematically evaluate single-point mutations and their combinations [12]. While these pioneering efforts laid the groundwork, modern strategies increasingly leverage machine learning (ML) and data-driven approaches to predict affinity or interaction energy and identify beneficial mutations. The input here is typically a co-crystal or co-folding structure of antibody-antigen with the model tasked in either predicting the energy of a set of mutations, or proposing a set of favorable ones (Figure 3B).

One can perform affinity maturation either in a supervised or unsupervised fashion. In the supervised variety, one relies on datasets of affinity maturation datasets such as SKEMPI or AB-BIND. Unsupervised methods learn a variety of proxies from larger datasets (e.g. rotamer density), that are assumed to influence binding energy, thus side-stepping the need for labeled data.

Supervised methods for antibody affinity maturation rely on training models on structural affinity datasets such as SKEMPI and AB-BIND, often supplemented with synthetic data to address limited experimental availability. Notable examples include MVSF-AB, which combines features from ProteinBERT embeddings and physicochemical properties through CNNs and MLPs, achieving state-of-the-art accuracy in predicting mutant and natural antibody-antigen binding affinities. Similarly, the Antibody Random Forest Classifier (AbRFC) integrates structural and mutational data to predict affinity-enhancing mutations, successfully designing SARS-CoV-2 antibodies with up to 1000-fold binding improvements against Omicron variants. Graphinity [61], an equivariant graph neural network, learns atomic-resolution interaction patterns and achieves Pearson correlations nearing 0.9 on affinity datasets, demonstrating strong generalization through the use of both experimental and synthetic ΔΔG data. However, the study also highlighted the need for tens to hundreds of thousands of high-quality experimental data points for fully generalizable predictions, reflecting current limitations in dataset size and diversity.

By contrast, unsupervised methods for antibody affinity maturation focus on learning from structural data without requiring labeled binding affinities, offering data-efficient alternatives to supervised approaches. Models like FvHallucinator [62] use generative hallucination to design sequences by minimizing geometric loss between predicted and target structures, successfully recovering native-like CDR sequences and generating functional binders validated via Rosetta. GearBind [63], a geometric graph neural network, combines large-scale pretraining on protein structures with fine-tuning on datasets like SKEMPI to predict mutations that significantly enhance affinity, achieving up to 17-fold improvements experimentally. DSMBind [64] employs energy-based modeling with SE(3) denoising score matching, learning to reconstruct perturbed structures and generating nanobody designs validated by ELISA assays, showcasing its versatility across binding tasks. Similarly, RDE-PPI [65] leverages a flow-based generative model to estimate rotamer probability distributions, using entropy to predict binding free energy changes (ΔΔG). Trained on structural data, it outperformed traditional methods on the SKEMPI2 dataset and successfully ranked affinity-enhancing mutations in a SARS-CoV-2 antibody design. Together, these methods highlight the potential of unsupervised learning to generate and optimize antibodies with minimal reliance on labeled data.

While these computational strategies excel at affinity maturation, they generally rely on the availability of high-resolution antibody-antigen complex structures, such as those derived from X-ray crystallography. This dependency poses a challenge, as generating accurate models of antibody-antigen complexes remains non-trivial. Advances in structure prediction methods for co-folding are increasingly addressing this bottleneck, aiming to expand the applicability of affinity maturation techniques even in cases where experimental structures are unavailable.

**Co-folding - structure-prediction-based design of binders**

Recent advances in structure prediction of monomers [15] have spurred an array of antibody variable region specific models [66]. Such models now make it possible to provide predictions that are of higher quality than previous homology models, typically in high-throughput and with low memory requirements [67]. For applications such as protein design, one would expect to model large numbers of variants, so the models have evolved to produce answers much faster than the pioneering AlphaFold-2 software, for which modeling even 1,000 antibodies would be cost and time prohibitive [68–70]. High throughput modeling of single structures is desirable for scaffold design and inverse folding that are covered in later sections.

Modeling of individual structures has naturally evolved into tackling multimeric complex prediction or 'co-folding' - akin to classic global protein docking. Development of AlphaFold2-multimer also started a trend of using the scores [71] from the models to additionally assess the quality of binding between an antibody and the antigen (Figure 3c). This is somewhat different from antibody-antigen docking, where one is interested in re-establishing the complex, but rather using the intermediary scores, such as iPTM+PTM to assess whether an arbitrary antibody (or protein) could bind to a given antigen as an oracle [72].

Recent studies have highlighted the application of advanced structural prediction methods in improving antibody-antigen docking and design. Yin and Pierce (2024) [73] evaluated AlphaFold2 (AF2) for refining docked antibody-antigen complexes by using stripped side-chain templates as input. AF2 improved docking performance, particularly in bound complexes, by retaining 50% of decoy contacts and refining interface structures with an average shift of 1.24Å, although its rescoring efficacy diminished with lower model quality. Wu et al. (2024) [74] introduced tfold-AB, a multi-task model leveraging AlphaFold2 and large language models for flexible docking and virtual screening, achieving DockQ scores of 0.217 in global and 0.416 in local docking scenarios. It showed potential for enriching antibody hits against targets like PD1 and SARS-CoV-2 antigens. Bang et al. (2024) [75] developed GaluxDesign, which achieved near-atomic accuracy (1.4Å RMSD) in challenging CDR-H3 loop predictions using inter-chain features and a novel G-pass scoring metric. The model outperformed AlphaFold and other tools in predicting HER2 binding, generating novel antibodies with high experimental success rates, including 13.2% for HER2-targeting designs.

Recently, diffusion-based improvements in AlphaFold3 were focused specifically on antibodies, improving the model performance on this modality upon AlphaFold2 [76]. Nevertheless, antibodies appear to be a particularly problematic format that still eludes such state-of the art modeling attempts. Currently there are community efforts to reproduce the successful architecture of AlphaFold3. Such reproductions, however, appear to be running into the same issues, indicating that global antibody-antigen docking/co-folding is still out of reach, and to get at reasonable models one needs to provide some constraining epitope information to the model [77].

Collectively, these advances demonstrate significant progress in antibody docking, rescoring, and de novo design, but one that still needs to reach a level that can be translated into clinical applications. Antibody CDRs are consistently eluding attempts to predict them accurately [78]. One confounding factor here might be CDR flexibility, as most of the methods treat 3D coordinates as static snapshots rather than means to an ensemble [79].

Given such shortcomings, predicting structures of an antibody-antigen complex can be seen as a proxy of assessing the viability of a given antibody sequence against an antigen. Exhaustive enumeration of such sequences is possible, but most of the structural methods above would make it computationally prohibitive to score. Such enumeration does not even take  For this reason designing an antibody given an antigen, the so-called 'de novo' design has always been of great interest.

**Sequence-structure co-design.**

The ultimate goal of structure-based antibody design is to develop a novel binder against a given epitope. Some of the early methods approached this without resorting to machine learning to assemble novel binding structures using fragments (Figure 3D). Examples here include OptCDR, AbDesign, RosettaAntibodyDesign and the method by Rangel et al.

Early computational methods for antibody design leveraged structural data to generate and optimize complementarity-determining regions (CDRs) for antigen binding. OptCDR [80] focused on canonical CDR structures, utilizing energy minimization and mutational libraries to design diverse CDRs, although its designs were not experimentally validated. AbDesign [81] used structural and sequence data from the Protein Data Bank (PDB) combined with Rosetta-based docking to optimize binding affinity and stability, successfully recapitulating natural backbone conformations in several benchmarks. RosettaAntibodyDesign (RAbD) [11] improved upon this by incorporating Monte Carlo minimization to graft and optimize CDRs, with experimental validation showing up to 12-fold improvements in binding affinity. More recently, fragment-based approaches like that of Rangel et al. (2022) [82] used structural fragments from PDB datasets to design single-domain antibodies with optimized stability and nanomolar affinities, validated against targets such as SARS-CoV-2. Together, these methods showcase the evolution of computational tools in antibody design, with increasing emphasis on experimental validation and real-world applicability.

Recent advances in antibody design have integrated machine learning with structural data, moving beyond traditional structure-based methods. RefineGNN [83] pioneered this approach by representing antibody sequences and structures as graphs, using message-passing networks to co-design complementarity-determining regions (CDRs) for improved binding affinity and neutralization. Trained on data from SAbDab and CoVAbDab, it showed strong performance in computational tasks like antigen-binding and SARS-CoV-2 neutralization but lacked experimental validation. Similarly, MEAN (Multi-channel Equivariant Attention Network) framed antibody design as a conditional graph translation problem, leveraging E(3)-equivariant message passing and attention mechanisms to predict CDR sequences and structures [84]. It outperformed baseline methods in computational benchmarks, including CDR-H3 design and binding affinity optimization, yet also lacked direct experimental testing. Building on this, dyMEAN [85] introduced full-atom modeling and the "shadow paratope" concept to better capture antigen-antibody interactions, further enhancing computational performance in structure prediction and affinity optimization. While these methods demonstrate promising results in silico, their lack of in vitro validation remains a limitation.

Recent advancements in computational antibody design have utilized diffusion-based generative models, which iteratively refine antibody sequences and structures by reversing noise corruption processes. DiffAb [86] was one of the first to apply this approach, using antibody-antigen complexes from SAbDab to co-generate CDR sequences and 3D structures. It demonstrated strong computational performance on targets such as SARS-CoV-2 and influenza but lacked experimental validation. Similarly, AbDiffuser [87] combined sequence-structure relationships with physics-informed constraints, achieving successful in vitro validation, with 37.5% of HER2-specific antibodies showing tight binding affinities comparable to Trastuzumab. AbX [88] extended diffusion modeling by integrating evolutionary, physical, and geometric constraints, leveraging pre-trained protein language models and structural data to optimize antibody-antigen binding. Though computationally robust, it also lacked experimental validation. Antibody-SGM [89] focused on heavy-chain design, using a score-based diffusion process to generate full-atom structures, further refined by Rosetta, and confirmed stability via molecular dynamics simulations. Despite promising computational results, it too remains unvalidated in wet-lab settings. Together, these diffusion-based methods showcase significant potential for antibody design but highlight a recurring gap in experimental confirmation.

All the methods described were solving a problem of combining the structure and sequence optimization. As the last graph and diffusion-based methods exemplify, the focus is shifting more towards machine learning generative methods. Though the methods covered here approaches simultaneous sequence-structure optimization there is a set of modern methods that split the problem in firstly generating the backbone, followed by predicting a sequence that could fit it.

**Scaffold design**

It is known that multiple protein sequences can adopt similar structures. This paradigm is exploited in 'scaffolding' which aims to generate novel protein backbones able to interact with another protein of choice (Figure 3E). Backbone generating methods are exemplified by methods such as IgDiff [90], Ig-VAE [91], and Sculptor [92].

IgDiff utilizes SE(3) diffusion to model antibody backbones and employs AbMPNN for sequence generation, trained on synthetic antibody structures from the Observed Antibody Space (OAS) and ABodyBuilder2 predictions [69]. Experimentally, 28 IgDiff designs showed high expression yields, validating its practical applicability. Ig-VAE, designed specifically for antibodies, generates 3D atomic coordinates for immunoglobulin domains using a rotationally and translationally invariant VAE, trained on AbDb/abYbank datasets. It demonstrated strong computational performance, such as epitope-specific SARS-CoV-2 RBD design, but lacked experimental validation. Sculptor, an evolution of Ig-VAE, integrates molecular dynamics simulations and interaction-guided modeling to design binders for user-specified epitopes. Combining VAE-generated backbones, sequence optimization, and Rosetta refinement, Sculptor successfully designed a broadly neutralizing binder for snake venom toxins, experimentally validated for cross-reactivity with multiple toxins. These methods highlight the growing potential of generative models in antibody and binder design, however as in many other cases, experimental validation was limited.

To the best of our knowledge, the current state of the art in backbone generation, which includes full experimental validation, was the fine-tuning of RFDiffusion [18]. It was trained on backbones with introduced noise, with a network tasked to recreate the original coordinates. In order to operate within the sphere of antibodies/nanobodies, RFDiffusion had to be fine-tuned on antibodies from the Protein Data Bank (PDB) [21]. This approach was developed using a combination of nanobody and general protein structures from the Protein Data Bank to train the model. The algorithm works by using a noising and de-noising process to iteratively refine protein backbones, specifically focusing on generating diverse CDR loop conformations and nanobody-antigen binding orientations. After designing the structures, ProteinMPNN is used to optimize the sequences of the CDR loops. The method was benchmarked both computationally and experimentally: it was applied to design nanobodies targeting a range of disease-relevant antigens (including influenza hemagglutinin, RSV, and SARS-CoV-2), and the resulting designs were experimentally validated through surface plasmon resonance (SPR) and cryo-electron microscopy (cryo-EM). The cryo-EM results confirmed that one of the designed nanobodies closely matched its predicted structure, validating the method's accuracy at atomic resolution. It should be noted that at the time of publication of this review, this promising method has not yet released any public tool associated with it.

For full protein/antibody design, the structural scaffolds generated need to be designed with sequences. This is the domain of inverse folding that models a sequence given a rigid structure. The scaffold generation and inverse folding algorithms are used currently in conjunction for a full protein/antibody design pipeline.

**Sequence design in structural context - inverse folding.**

It is often desirable to improve upon a known protein sequence, and the task is made easier if its coordinates are known. Altogether the problem is known as 'inverse folding' (Figure 3F) for its clear shift in the prediction objective to the protein folding problem. Though the problem has been known and tackled for a very long time [93], recent advancements in protein structure prediction and antibody sequence generation have resulted in a revival of such methods. Millions of predicted structures by AlphaFold2 can be used to train inverse folding algorithms. Large-scale generation of antibody sequences allows us to model these and develop antibody-specific antibody inverse folding methods.

Protein-generic inverse folding methods generally form the foundation for antibody-specific design approaches, with ProteinMPNN and ESM-IF being two prominent examples. ProteinMPNN [19] uses a message-passing neural network (MPNN) to predict sequences that fold into given protein structures by encoding features like atomic distances and frame orientations. It achieves high sequence recovery rates and structural fidelity, outperforming traditional methods such as Rosetta. ProteinMPNN has been validated both computationally and experimentally, with techniques like X-ray crystallography and cryoEM confirming its ability to accurately fold into target structures. ESM-IF [20,22] leverages a GVP-Transformer model trained on a dataset of 16,000 experimental and 12 million AlphaFold2-predicted structures, achieving notable improvements in sequence recovery, particularly for buried residues. It demonstrated its utility in practical applications by introducing point mutations to anti-SARS-CoV-2 antibodies [22], which enhanced binding affinity in experimental validations.

ESM-IF and ProteinMPNN were trained on a large corpus of proteins, but arguably a very small sample of all allowed antibody structures. To the best of our knowledge, two antibody-specific inverse folding methods were developed, AbMPNN and AntiFold, fine-tuning ProteinMPNN and ESM-IF respectively on a modeled antibody corpus.

AbMPNN [94], trained on 3,500 antigen-binding fragments from SAbDab and 147,919 paired variable regions from OAS using ABodyBuilder2-derived structures, achieved 60% sequence recovery for CDR loops—outperforming ProteinMPNN's 40%—and showed a 20% improvement in median RMSD for CDR-H3 loops, enhancing designability and stability. AntiFold [95], trained on 2,074 experimentally solved and 147,458 predicted antibody structures, excelled in amino acid recovery (60% for CDR-H3) and achieved a Spearman's rank correlation of 0.418 for antibody-antigen binding affinity, surpassing AbMPNN and ESM-2. IgDesign [96] focused on designing complementarity-determining regions (CDRs) for eight therapeutic antigens, generating 1 million sequences per antigen and filtering them for in vitro testing. It achieved superior binding rates across antigens, with statistically significant improvements for 7 out of 8 HCDR3 targets, making it a standout for experimental validation. While IgDesign demonstrated experimental success, it is not freely available, unlike AbMPNN and AntiFold that are free, but did not demonstrate experimental validation.

Inverse folding methods represent a powerful alternative to simultaneous sequence-backbone design by assuming that the fold will not change. This is oftentimes desirable as structure is crucial to antibody-antigen recognition and only minute changes need to be introduced, maintaining the fold, but modulating the overall function of the antibody.

**Generating antibodies with improved functions - developability optimization.**

Most of the review, and much of antibody design is focused on binder development. However developing a binder is arguably an experimentally solved problem with the bigger wet-lab hurdle being the optimization of subsequent developability properties. The iterative process of fine-tuning the myriad biophysical properties is not linear and can account for much of the time and effort in the preclinical stage [42,44,97].

Antibody design methods that address developability issues are much more heterogeneous both in their approaches and goals. Arguably, binder development has one objective, which is generally a high affinity interaction. A single developability property on the other hand, such as self-association, can have several assays associated with it that might not be directly comparable to one another. There is also great scarcity of data on developability points. To date, the Jain characterization of ca.

100 therapeutic antibodies, remains one of the most comprehensive characterizations of developable antibodies - however without negative data points [98].

Because of data scarcity, many methods, such as the Oracles (Figure 3A), focus on close-to-exhaustive enumeration of binders, followed by computational filtering for developable antibodies using general tools such as the Therapeutic Antibody Profiler [99] or Camsol [100].

There also exists a plethora of work dedicated to the computational prediction of individual assay data and properties that can lead to more developable models and be incorporated in generative protein design models. A full review of these models is beyond the scope of this review, but generally these include Hydrophobic Interaction Chromatography (HIC), expression (concentration, purity), stability, chemical modifications (D isomerization, N deamidation), enzymatic PTMs including phosphorylation and glycosylation, immunogenicity (T-cell epitope), Pk properties such as clearance, and viscosity. Many of the published methods for these properties are not commercially or academically available as they are developed using scarce proprietary data, however, with the advent and broad participation of consortia such as the FAITE consortium (https://faiteconsortium.org/), this could be shifting.

An alternative to enumeration followed by binder validation is biased generation of sequences through multi-property optimization (MPO). Amimeur et al. (2020) [101] pioneered this approach using a Generative Adversarial Network (GAN) trained on over 400,000 sequences from the Observed Antibody Space (OAS) to produce "humanoid" antibodies with human-like structural and functional diversity. The GAN, fine-tuned for therapeutic traits such as stability and low immunogenicity, was experimentally validated through assays like DSF, SEC, and SINS, achieving a high success rate in generating antibodies with desirable developability profiles. Building on this, Turnbull et al. (2024) [102] introduced p-IgGen, a GPT-2-based model fine-tuned on paired and developable antibody sequences, which excelled in immunogenicity prediction while maintaining computational efficiency. Complementing these generative approaches, Hutchins et al. (2024) [103] used the DeepAb model to design 200 anti-HEL antibody variants, optimizing thermostability and affinity through mutations informed by deep mutational scanning and achieving significant experimental success, with most variants showing increased stability and up to a 21-fold affinity improvement. Dreyer et al. [104] extended this pipeline for one-shot antibody discovery, using computational tools like AbMPNN and ESM to design SARS-CoV-2 RBD-binding antibodies with enhanced developability, validated experimentally through stability and aggregation assays, achieving a 54% success rate against escape mutations.

Altogether, though there appears to be progress in generating and designing antibodies with improved developability properties, most efforts are one off proofs of concept. To close the gap between experimentation and these methods being employed to develop novel drugs, data generation, method integration and benchmarking is necessary.

**Outstanding challenges - data & experimental validation.**

The biggest issue within the protein design field remains not model development but data availability, both for training and benchmarking. Unlike in text or image generation fields, where it is fairly cheap to gather datasets of millions of data points, in biology it is not. Data generation is prohibitively expensive with large discrepancies between cheaper but less informative (sequences) and more expensive but more informative datasets (e.g. structures).

There are only ca. 15,000 non-redundant single chain protein structures in the protein data bank [28] - compared to several billion sequenced chains [105]. On the antibody front, there are several billion sequences available [106–108], but only several thousand non-redundant structures [49]. The number of non-redundant antibody-antigen complexes is smaller still, being in the order of around 1000 structures.

The structure datasets are crucial to the development of backbone generation protocols, such as RFDiffusion. Paucity of such data, especially on the antibody-antigen complex front is a blocker for development of better algorithms, as even the latest iteration of AlphaFold, falls short of providing an actionable solution to the antibody-antigen interaction problem. The structural antibody datasets such as SABDAB [49] and ABDB [109] have become proxies for benchmarks as they compile antibody-antigen information specific to antibodies. For the design tasks, the complexes are employed to train backbone generation algorithms. Much more specialized datasets that also gather affinity information are SKEMPI [110] and AB-BIND [111]. Here, the structure of an antibody-antigen is accompanied by mutation and affinity measurement. Nevertheless, these datasets contain a small number of structures and measurements relative to the scale of the problem (several hundred data points each).

The copious sequence datasets are employed to learn meaningful representations of molecules, given the unavailability of the corresponding large-scale structure data. For structure-informed sequence design using inverse folding, the paucity of structural data is side-stepped by model generation. For instance ESM-IF was trained using 12m AlphaFold models. In the antibody space, the widely used resource is the

Observed Antibody Space [107,108], which curates repertoire data used chiefly for language model training. Similar to ESM-IF training, OAS data were modeled and used for training the ABMPNN and AntiFold, to extend the datasets beyond the several thousand available structures. Though it is plausible to re-use such experimentally generated data, care needs to be exercised as it has been raised that such data might in fact be biasing the models in the undesirable direction [112]. The fact remains that though there is a lot of sequence data, hardly any of it is associated with binding or developability data points.

The paucity of data highlights another large problem, which is experimental validation. Many design methods are proposed purely in silico, without experimental validation. Benchmarking is done on publicly available datasets - which have few data points for structures (~1000), affinity (100s) and even less for developability (10s, depending on which assay one focuses on [113]. Because of the paucity of the data, generative methods are often compared on pre-existing datasets [114]. In an ideal scenario, the generated sequences would in each case be made in the lab and the structures solved, as is done in few cases.

Experimental validation, however, is very expensive and benchmarks on par with CASP or CACHE are only very recently coming into existence in the protein design world [115–117]. In the antibody-world, such benchmarking needs to take into account not only whether one can develop a binder but also how developable such sequences are [113]. Arguably, developing a binder is an experimentally solved problem, with the biggest remaining issues being hitting the right epitope, specificity, and developability. Furthermore, any kind of benchmarking needs to weigh the speed and cost of computation versus purely experimental discovery. For this reason, the antibody-specific benchmarking that comes into existence takes this into account through both developability challenges as well as carefully weighing the benefits of experimental versus purely computational approaches [118,119].

Altogether, protein and antibody design fields are still in their infancy. It will take time, both in development and adoption before computational methods become the driver of discovery and design. Given the state of the field, we think that the change will be fuelled chiefly by creating large complex datasets adorned with developability data together with conscientious benchmarking and collaboration across industry and academia.

**Table 1. Methods in antibody design.** Methods are broadly categorized based on their inputs, antibody/antigen specific focus and the role they play in the design pipeline (end-to-end or just providing sequence for backbone). In terms of benchmarking, we indicate the extent of experimental validation. Otherwise there appears to be no single metric of success amongst the methods. The list is not comprehensive as it is intended to demonstrate the methods associated with their respective categories.

| Category | Method | Model architecture | Structure aware | Antibody specific | Antigen condition | Performance verification | Experiment verified | Availability | Citation |
|---|---|---|---|---|---|---|---|---|---|
| Zero-shot, few-shot sequence design | ESM-1B | Transformer architecture | No | No | No | This model outperformed sequence-based baselines models e.g. AbLang, abYsis, UniRef90 or Sapiens. It has a higher fraction with improved binding, median fold improvement and maximum fold improvement for selected antibodies. | Yes (Model was evaluated through experimental assays measuring the binding affinities of antibodies with model-predicted mutations. Model-guided mutations resulted in significant improvements in binding affinity) | https://github.com/facebookresearch/esm | 23 |
| | ProseLM | Based on ProGen2 (autoregressive transformer), a pre-trained protein language model. Model uses Message Passing Neural Networks (MPNN) and invariant-point message-passing (IPMP) layers | Yes | No, but with additional fine-tuning for specific antibodies using data from the Structural Antibody Database (SAbDab) | No, but it was tested on antibody-antigen complexes | proseLM-XL achieved higher recovery rate for native sequences reaching 3.59% higher median recovery rate than the causal encoder. The ProseLM models trained with coordinate noise achieved higher rates of single-sequence prediction structure prediction success with AlphaFold2 and yielded more confident structures in comparison with ProteinMPNN | Yes | https://github.com/P326rofluent-AI/proseLM-public | 48 |
| | Shanehsazzadeh et al. | Zero-shot generative design approach. It includes two step process: MaskedDesign (3D backbone structure of a bound antibody-antigen complex prediction) and IgMPNN (HCDR sequences prediction) | No | Yes | Yes (generating binders specific to HER2, VEGF-A, and the SARS-CoV-2 spike protein, and the designs were validated for binding to these antigens.) | The generative AI model achieved top 1,000 binding rates of 10.6% (HCDR3) and 1.8% (HCDR123), outperforming baselines by 4- to 11-fold, while off-target designs showed a 3-fold drop, highlighting | Yes (The model was validated experimentally. Out of 440,000 generated HCDR3 variants, approximately 4,000 were estimated to bind to HER2 based on screening, with 421 | Not mentioned, only HER2 binders and measured binding affinities are open-sourced at available at https://github.com/AbsciBio/unlocking-de-novo-antibody-design | 50 |

| | | | | | | its antigen-specific accuracy. | confirmed binders validated through SPR (Surface Plasmon Resonance) | | |
|---|---|---|---|---|---|---|---|---|---|
| Oracles | Mason et al. | Convolutional Neural Network (CNN) | No | Yes | Yes, trained on known binders and non-binders. | The model achieved an area under the ROC curve (AUC) of 0.91 and an average precision of 0.83. 30/30 experimentally validated variants retaining antigen specificity and 85% exhibiting nanomolar affinity | Yes (The model was experimentally validated by expressing selected antibody variants in mammalian cells. Results showed that the majority of predicted binders demonstrated specific binding to HER2) | https://github.com/dahjan/DMS_opt | 38 |
| | Lim et al. | Convolutional Neural Networks (CNN) to classify antibody binders and non-binders, and Generative Adversarial Networks (GANs) to generate synthetic antibodies | No | Yes | Yes (trained on known binders and non-binders to the specific antigens PD-1 and CTLA-4) | The CTLA-4 and PD-1 models achieved respectively: prediction accuracy of 91.2% and 92.6%, Matthews correlation coefficient (MCC) of 0.68 and 0.78, areas under the curve (AUC) of the receiver operating characteristic (ROC) 0.9 and 0.94. | Yes (The model was experimentally verified by testing generated antibody sequences against the antigens PD-1 and CTLA-4) | https://github.com/ywlim/Antibody_deep_learning | 54 |
| | Chinery et al. | Convolutional Neural Network (CNN), Equivariant Graph Neural Networks (EGNN) and FLAML | No | Yes | Trained on binders and non-binders to HER2 | The model achieved a PR AUC of 0.71 with 170 training sequences, increasing to 0.94 with 28,900 sequences. Computational library design predicted binder enrichments of 19–30% and the experimental verification is ongoing. | Yes (Binding predictions were tested through Biolayer Interferometry, with further validation ongoing to assess the binding properties of top designs) | https://github.com/oxpig/Tz_her2_affinity_and_beyond | 39 |
| | Ursu et al. | Neural network called SN10 | No | Yes (trained on CDRH3) | The model is trained on specific antigens but does not condition the model on antigen properties for each task | The model achieved ID accuracy of 85-99% and OOD accuracy of 90-96% in challenging antigen-specific tasks, outperforming baseline methods. Compared to logistic regression, it demonstrated a 12% improvement in challenging scenarios, | The model was not directly validated in a wet lab; however, its experimental validation relied on data from 24,790 CDRH3 sequences binding to the HER2 antigen, derived from high-throughput | https://github.com/csi-greifflab/negative-class-optimization. | 120 |
| | | | | | showcasing superior generalization and binding rule discover | sequencing and affinity screening methods. This provided an indirect but robust evaluation of its performance against real-world binding data | | | |
| | AlphaBind | Transformer encoder with 4 heads and 7 layers. Antibody and target are encoded using ESM-2nv | No | Yes | Yes (tested on multiple antibody-antigen systems such as Pembrolizumab-scFv (targeting PD-1) and VHH72 (targeting SARS-CoV-2 RBD)) | The AlphaBind model demonstrated strong computational verification across multiple benchmarks. It achieved up to 74x affinity improvement in silico (for the best AAB-PP489 variant), with predicted affinities validated experimentally for 15/15 (100%) of the top candidates, all outperforming their parental antibodies. | Yes (The model's performance was verified using experimental methods. In vitro validation showed that AlphaBind-derived candidates consistently outperformed parental antibodies in binding affinity. For example, the top candidate for AAB-PP489 achieved a 74x improvement in affinity, and VHH72 candidates showed up to a 14x improvement) | https://github.com/A-Alpha-Bio/alphabind | 60 |
| Affinity maturation | Fv Hallucinator | Structure-conditioned design framework that utilizes pretrained deep learning models, specifically the DeepAb model (1D ResNet (1D convolution followed by three 1D ResNet blocks) and the bi-LSTM encode) | Yes | Yes | Yes | The model achieved over 50% sequence recovery for CDRs with wildtype seeding, designed human-like interfaces, and retained target conformation in 70% of designs (RMSD ≤ 2.0 Å). Additionally, 27% of designs showed improved binding energies compared to the wildtype | Yes (The model has been experimentally validated by comparing generated libraries with known experimental libraries and performing binding affinity tests, particularly in applications such as HER2-specific binding) | https://github.com/RosettaCommons/FvHallucinator | 62 |
| | AbRFC | Random Forest classifier with features engineered from structural and biophysical data, including metrics from Rosetta software and previously validated | Yes | Yes | Yes (it has been trained and validated on antibody-antigen complexes to predict mutations that retain or improve binding | The model achieved up to 1000-fold affinity improvement in two rounds of testing with <100 designs per round and outperformed alternatives like GNN and | Yes (The model predicted mutations were tested in a wet lab, confirming enhanced binding affinities in two rounds of screening, which yielded | https://github.com/tbc01/AbRFC | 121 |

| | | | | | | | | |
|---|---|---|---|---|---|---|---|---|
| | | metrics like AIF and SIN scores | | | affinity) | LLM models, achieving average PR AUC of 0.87 and identifying 22–31% affinity-enhancing mutations in tested datasets | optimized antibody variants with significantly improved binding) | | |
| | GearBind | Geometric graph neural network (GNN) that uses multi-level geometric message passing with pretraining with contrastive learning on large structural datasets to enhance its effectiveness | Yes | No (but it was tested specifically on antibodies for affinity maturation) | Yes (applied and verified on specific antigens. It was tested with antibodies against SARS-CoV-2 and the oncofetal antigen 5T4) | The model achieved Pearson R = 0.676 and Spearman R = 0.525 on SKEMPI, outperforming FoldX (Pearson R = 0.491) and Bind-ddG (Spearman R = 0.443). On 419 HER2 variants, GearBind achieved the highest correlations among models with Pearson R = 0.707, compared to FoldX (Pearson R = 0.411). | Yes | https://github.com/DeepGraphLearning/GearBind | 63 |
| | DSMBind | Energy-based model (EBM) that uses SE(3)-invariant neural networks. It also includes frame-averaging neural network and SE(3) denoising score matching (DSM) | Yes | No | Yes (For antibody-antigen complexes, the input includes CDR regions and antigen epitopes to evaluate binding energies.) | The model achieved strong computational performance, with an Rs = 0.374 for antibody-antigen binding and Rs = 0.388 for protein-ligand binding, outperforming other models. For protein-protein interactions, it achieved Rs = 0.403 in mutation effect predictions. | Yes | https://github.com/wengong-jin/DSMBind | 64 |
| | MVSF-Ab | Multi-view sequence feature learning framework combining semantic and residue features. It employs a pre-trained ProteinBERT model for embedding antibody and antigen sequences, followed by a Convolutional Neural Network (CNN) for semantic feature extraction and a multilayer | No | Yes | No (but it is evaluated on datasets containing various antibody-antigen pairs) | The model was verified on multiple datasets, achieving Root Mean Square Error (RMSE) and Pearson Correlation metrics on datasets such as SAbDab, AB-Bind, and SKEMPI 2.0. For example, on SKEMPI 2.0, it achieved an RMSE of 1.513 and a Pearson correlation of 0.671, indicating robust predictive capability | No | https://github.com/TAI-Medical-Lab/MVSF-AB | 122 |

| | | | | | | | | |
|---|---|---|---|---|---|---|---|---|
| | | perceptron (MLP) for residue-based features derived from AAindex | | | | compared to other methods. | | | |
| | RDE-PPI | flow-based generative model. The method integrates a Rotamer Density Estimator (RDE) for modeling sidechain conformations, an entropy-based algorithm for estimating conformational flexibility, and neural networks to predict changes in binding free energy (ΔΔG) with high accuracy. | Yes | No (but it has applications for antibodies e.g., optimizing CDRs in antibody-antigen interactions) | Estimates the rotamer density so it does include the partner binder information to a certain extent | Computationally, RDE-Network achieved superior performance in the change of binding free energy prediction compared to baselines like Rosetta and FoldX, with a Pearson correlation of 0.6447 and Spearman correlation of 0.5584 on the SKEMPI2 dataset. It ranked favorable mutations for SARS-CoV-2 antibodies effectively, placing three beneficial mutations in the top 10% of predictions. | No | https://github.com/luost26/RDE-PPI | 65 |
| Co-folding | Galux Design | Structure-conditioned model focusing on antibody loop (especially H3 loop) structure prediction and generation. It integrates components like AlphaFold-Multimer for structural templates and a proprietary scoring mechanism based on structural confidence metrics | Yes | Yes | Yes | The model achieved top H3 loop prediction accuracy (1.4 Å RMSD on GAbD), outperforming AlphaFold (2.4 Å) and ABlooper (3.4 Å). It excelled in antibody loop design with a 26% G-pass rate and 55% structure recovery. In vitro, it showed high success rates for PD-L1 (15%), PD-1 (5–9%), and EGFR mutants (8%), with sub-nanomolar affinities and strong specificity | Yes (The model's predictions were experimentally validated through in vitro binding assays. Success rates for binding antibodies exceeded those of previous designs, with a success rate of up to 13% for HER2 loop designs and sub-nanomolar affinities confirmed for several targets) | Not mentioned | 75 |
| | Tfold-ab | Based on AlphaFold2, using language models in the place of Evoformer | Yes | Yes | Yes | Performance was validated on the IgFold-Ab and SAbDab-22H1-Ab benchmarks, where it achieved the lowest RMSD values in CDR regions compared to other methods, with a 2.74Å RMSD on the CDR-H3 region. It also performed well on orientational | No | https://drug.ai.tencent.com/en | 74 |

| | Model | Architecture | Open source | Antibody specific | Incorporation of structural constraints | Performance | Web interface | Link | Ref |
|---|---|---|---|---|---|---|---|---|---|
| | | | | | | metrics, demonstrating strong accuracy in antibody structure prediction. | | | |
| | AlphaFold2, AlphaFold multimer | Based on attention mechanisms. AlphaFold utilizes end-to-end neural networks, using AlphaFold-Multimer. | Yes | No (but has been applied and benchmarked extensively on antibody-antigen complexes) | AlphaFold (v.2.0)- no AlphaFold-Multimer (released in AlphaFold v.2.1) - yes | AlphaFold v2.2 achieved acceptable or higher accuracy in 26% of antibody-antigen complexes (n=427), with medium or higher accuracy in 18% and high accuracy in 5%; these rates increased to 37%, 22%, and 6%, respectively, when considering all 25 predictions per complex. The updated v2.3 model improved medium or higher accuracy to 36% of top-ranked predictions, and the AFsample protocol further enhanced it to 51%, demonstrating notable advancements in predictive performance. | No | https://github.com/piercelab/alphafold_v2.2_customize and AlphaFold2.2, AlphaFold2.3, and Colab-Fold antibody–antigen models generated in this study are available at https://piercelab.ibbr.umd.edu/af_abag_benchmarking.html | 73 |
| | AlphaFold3 | Diffusion-based generative architecture with simplified Pairformer module, a diffusion module, confidence modules and trunk network. | Yes | No (but it significantly improves accuracy for antibody-antigen complexes and is validated for protein-protein interactions, including antibodies) | Yes ( can be antigen-conditioned. It predicts antibody-antigen interactions by incorporating epitope structures and optimizing CDR regions for binding specificity ) | Yes, it demonstrated increased accuracy in comparison with AlphaFold-multimer v2.3 e.g. Protein-antibody DockQ scores (a measure of interaction quality) increased significantly with top-ranked predictions exceeding 80% accuracy in certain benchmarks., improved success rates for antibody-antigen predictions using 1,000 model seeds. | No | Not mentioned | 76 |
| | Chai-1 | Reproduction of Alphafold3 | Yes | No | Can be antigen-conditioned. It supports predictions constrained by experimen | Achieved a DockQ success rate on antibody-protein interactions : of 52.9% with MSAs( significantly | No | Available for non-commercial use https://github.com/chaidiscovery/chai-lab/. Authors also provide a web interface for commercial | 77 |
| | | | | | tally derived features, such as antigen epitope residues, to enhance antibody-antigen interface modeling | outperforming baseline models), in single-sequence mode, Chai-1 achieved 47.9% (surpassing AlphaFold 2.3's performance with MSAs) | | applications available at https://lab.chaidiscovery.com | |
| | Boltz-1 | Reproduction of Alphafold3 | Yes | No | No | Boltz-1 achieves a median LDDT of 0.54 and TM score of 0.31 for CASP15 RNA targets, compared to Chai-1's 0.41 and 0.31. On the curated PDB test set, both models perform similarly with DockQ > 0.23 and median TM scores. | No | Available for non-commercial use. https://github.com/jwohlwend/boltz | 123 |
| Structure sequence co-design | OptCDR | Computational workflow based on canonical structures for antibody complementarity-determining regions (CDRs). Model is implementing 4 step structure: selection of CDR canonical structures (backbone only), amino acid sequence initialization using mixed-integer linear programming (MILP), modified version of the previously developed iterative protein redesign and optimization (IPRO) and library generation | Yes | Yes | Yes (evaluates designs against specific antigens, such as hepatitis C capsid peptide, fluorescein, and vascular endothelial growth factor (VEGF)) | The model excelled in de novo antibody design, significantly improving binding metrics, such as interaction energy for a hepatitis C peptide (-62.6 to -175.8 kcal/mol) and increasing polar contacts (8 to 31). It also generated novel CDRs for VEGF and fluorescein with binding performance comparable to experimentally optimized antibodies. | No | https://www.maranasgroup.com/software.htm | 80 |
| | OptMAVEN | The model implements a computational framework that uses a modular approach for designing antibody variable regions. It employs a combinatorial optimization strategy to select and assemble CDRs from a pre-built database based on their compatibility and binding energy with the target | Yes | Yes | Yes (It evaluates the designed antibodies for binding affinity and specificity against specific target antigens using computational metrics.) | The OptMAVEn model achieved a 96% success rate in antigen positioning across 120 complexes and rediscovered 57.5% of native antibody parts during modular part selection. In the designed sequences 35% and 20% of mutations in the native AM influenza and HIV-1 | No | https://www.maranasgroup.com/software.htm | 124 |

| | | | | | | | | |
|---|---|---|---|---|---|---|---|---|
| | | antigen. | | | | antibody models, respectively, were recaptured. | | | |
| | AbDesign | Combinatorial backbone and sequence optimization algorithm. It leverages the Rosetta macromolecular modeling suite to design antibodies. It uses fuzzy-logic design for optimizing both ligand binding and antibody stability | Yes | Yes | Yes (the model has been used to design antibodies targeting specific antigens, including lysozyme, sonic hedgehog protein, and tissue factor. Designs are computationally docked and scored for binding to specific epitopes) | The model's computational verification demonstrates robust performance. It achieved >30% sequence identity with natural antibodies in 5/9 cases and backbone conformations within 1 Å RMSD for 4 designs, verifying structural similarity to natural antibodies. | No | The methods have been implemented within the Rosetta macromolecular modeling software suite54 and are available through the Rosetta Commons agreement, For additional information regarding RosettaScripts and implementation please see the RosettaScripts documentation page on the RosettaCommons website. (https://www.rosettacommons.org/manuals/archive/rosetta3.3_user_guide/RosettaScripts_Documentation.html) | 81 |
| | RAbD | Built on the Rosetta software suite. It employs a Monte Carlo + minimization framework for optimizing antibody sequences and structures. | Yes | Yes | Yes (It uses antigen-antibody complex data to evaluate binding interfaces and optimize designs for specific antigen-binding interaction) | The RAbD model was computationally verified using 60 antigen-antibody complexes with 6,000 design cycles per CDR. It achieved Design Risk Ratios (DRRs) of 2.4–4.0x for CDR recovery and an Antigen Risk Ratio (ARR) of 1.5 for antigen-contacting residues. | Yes | RAbD is publicly available as part of the Rosetta software suite. The necessary databases and tools can be accessed via https://rosettacommons.org/ and updated structural data from PyIgClassify. | 11 |
| | Rangel et al. | The model employs a fragment-based computational approach for designing antibody complementarity-determining region (CDR) loops. The strategy combines structural fragments (CDR-like fragments) and sequence information from databases like the Protein Data Bank (PDB) to create CDR loops optimized for binding specific epitopes. | Yes | Yes | Yes (validated for specific antigens, such as human serum albumin (HSA) and the SARS-CoV-2 spike protein receptor-binding domain (RBD)) | Computationally, the method targeted 78% of antigen surfaces with a density of 19.2 CDR designs per nm². When tested with both experimental and AlphaFold-predicted structures, 77% of designed CDRs were identical between models and crystal structures, confirming reliability across structure quality. | Yes | Not mentioned | 82 |
| | RefineGNN | Graph neural network | Yes | Yes | The model optimizes | The performance | No | https://github.com/ | 83 |
| | | (GNN) with generative capabilities | | | antibodies for antigen binding through benchmarks like amino acid recovery (AAR) for antigen-binding tasks. However, direct antigen structural conditioning is not included in its generation steps | was verified computationally. The model results include for instance 30% improvement in root mean square deviation (RMSD) for CDR-H3 structure prediction compared to AR-GNN or higher AAR (35.37%) on antigen-binding tasks, outperforming baselines like RAbD | | wengong-jin/RefineGNN | |
| | MEAN | E(3)-equivariant graph neural networks with alternating internal and external encoders and a novel attention mechanism to model 3D geometry and interactions within antibody-antigen complexes | Yes | Yes | Yes | It significantly outperforms baselines in 1D sequence and 3D structure modeling, achieving up to 36% improvement in amino acid recovery and lower RMSD values. In antigen-binding CDR-H3 design, MEAN achieves nearly perfect structural alignment with TM-scores exceeding 0.98 and RMSD as low as 1.81. Additionally, in affinity optimization, it outperforms previous methods by achieving the most substantial binding affinity improvements (ΔΔG of −5.33 kcal/mol), showcasing its ability to generate antibodies with high specificity and affinity. | No | https://github.com/THUNLP-MT/MEAN | 84 |
| | dyMEAN | Similar to MEAN | Yes | Yes | Yes | It achieves superior results in CDR-H3 generation with a 43.65% amino acid recovery (AAR), TM-score of 0.9726, and RMSD of 8.11, while excelling in docking quality (DockQ: 0.409) and affinity optimization with the most significant | No | https://%20github.com/THUNLP-MT/dyMEAN | 85 |

| | | | | | | binding affinity improvement (ΔΔG: -7.31 kcal/mol) | | | |
|---|---|---|---|---|---|---|---|---|---|
| | Diffab | Diffusion-based generative model that combines probabilistic modeling and equivariant neural networks. | Yes | Yes | Yes (It explicitly incorporates antigen structures into its predictions, allowing the generated CDRs to adapt to specific antigen binding sites.) | The model achieved up to 87.83% amino acid recovery (AAR) for CDR-H1, outperforming RAbD (65.75%) and FixBB (37.14%), with RMSD ≤ 1.5 Å for most CDRs except CDR-H3, which had RMSD up to 3.597 Å due to its structural diversity. For CDR-H3 optimization, it improved binding energy (IMP up to 23.63%) | No | https://github.com/luost26/diffab | 86 |
| | AbDiffuser | Denoising diffusion-based generative model specifically designed for antibody sequence and structure generation. The core of its architecture includes the Aligned Protein Mixer (APMixer), which is an SE(3) equivariant neural network. | Yes | Yes | Yes (incorporates antigen verification in its experimental setups. It was validated for specific antigens, such as HER2, showing successful generation of high-affinity binders) | The AbDiffuser model achieved 22.2% binding (raw) and 57.1% (filtered), with an average pKD of 8.70 and a best binder pKD of 9.50, surpassing Trastuzumab (pKD ~9.21). It required only 16 samples, demonstrating 26x greater efficiency, and generated structures with an RMSD of 0.4962, closely matching test sets. | Yes | Not mentioned | 87 |
| | AbX | Model employs a score-based diffusion framework with an ESM-2-guided encoder, Invariant Point Attention layers, and a recycling mechanism to co-generate antibody sequences and SE(3) structures | Yes | Yes | Yes | Outperformed baseline models (DiffAb, dyMEAN) across metrics like Amino Acid Recovery (AAR) and RMSD. For instance, it achieved 30.8% Loop AAR and 3.24 Å Loop RMSD on the RAbD test set, which are significant improvements. Demonstrated 18.64% Improvement Percentage (IMP) in binding energy for designed antibodies compared to natural counterparts | No | https://github.com/zhanghaicang/carbonmatrix_public | 88 |
| | Antibody SGM | Score-based UNet model that co-generates antibody heavy-chain sequences and structures using one-hot encoding and 6D inter-residue coordinates, refined with Rosetta | Yes | Yes | Yes | The model outperformed DiffAb in sequence recovery and RMSD for antigen-specific CDR generation achieving lower RMSD values (e.g., 0.818 Å for CDR-H1 and 2.901 Å for CDR-H3). Generated antibodies showed structural and sequence consistency with training data, achieving <1.5 Å RMSD for Rosetta-refined structures and similarity scores >65% with training datasets | No | https://github.com/xxiexuezhi/ABSGM | 89 |
| Scaffold generation | RFDiffusion | Based on RFdiffusion, uses an AlphaFold2/RoseTTAFold (RF2) inspired framework that applies Gaussian noise and iterative de-noising to predict protein backbones and antibody structure | Yes | Yes (fine-tuned specifically for antibodies, particularly single-domain VHHs and scFvs) | Yes | The model successfully designed VHHs with experimentally validated affinities ranging from 78 nM to 5.5 µM, achieving structural accuracy with backbone RMSD of 1.45 Å and CDR3 RMSD of 0.84 Å. | Yes (Designed VHHs were tested against specific disease-relevant targets (e.g., influenza HA, RSV, COVID-19 RBD) using biochemical assays, including surface plasmon resonance (SPR) and cryo-electron microscopy (cryo-EM)) | Not mentioned | 21 |
| | IgDiff | SE(3) diffusion-based generative model with Riemannian score-based generative modeling used for the diffusion of protein backbones. Sequences are predicted using the antibody-specific inverse folding model AbMPNN. | Yes | Yes | Yes (IgDiff can be antigen-conditioned for certain tasks. It enables design tasks where specific CDR loops or regions are generated while preserving the binding interface to target specific antigens) | The model generates antibodies with high consistency, achieving a self-consistency RMSD below 2 Å for all designs, with 88% meeting this threshold independently across all CDR loops. Compared to RFDiffusion, IgDiff excels in tasks like CDR H3 length changes, with 74% passing quality metrics versus 6% | Yes (Generated antibodies were made in the lab to verify that they can be produced) | https://zenodo.org/record/11184374 | 90 |
| | Sculptor | Variational autoencoder (VAE) for generating protein backbones tailored to specific epitopes | Yes | No | It generates binders specific to target epitopes, such as venom toxins and SARS-CoV-2 RBD, | The model achieved 1.2 Å RMSD in recovering native complex backbones, improving to 0.97 Å after refinement, with 68 | Yes (The model was experimentally verified. Designed proteins were expressed and tested for binding | Not mentioned | 92 |
| | | | | | and verifies designs using docking simulations and interaction energy evaluations. | interfaces outperforming native binders in ΔΔG. | using yeast surface display and fluorescence-activated cell sorting (FACS)) | | |
| | Ig-VAE | Variational autoencoder (VAE). It predicts 3D coordinates using a torsion- and distance-aware architecture | Yes | Yes | The model can be guided towards desired structural features through techniques like latent space sampling and constrained optimization to generate structures, such as SARS-CoV-2 RBD binders with high ACE2 epitope complementarity. | Model reconstructed torsion angles within ~10° and bond lengths within ~0.1 Å of experimental data, generated stable Ig backbones with low Rosetta energy scores, and achieved binding energies of ΔΔG e.g. -37.6 and -53.1 Rosetta units for SARS-CoV-2 RBD binders | No | https://github.com/ProteinDesignLab/IgVAE | 91 |
| Sequence Design with structure, inverse folding | ESM-IF | The model includes three main architectures: GVP-GNN, GVP-GNN-large, and GVP-Transforme | Yes | No | No | The GVP-Transformer achieved 51.6% sequence recovery, an 8.9% improvement over experimental-data-only models, maintained low perplexity (~4.01), and excelled in tasks like binding affinity prediction (Spearman 0.69) and zero-shot mutation effects | No | https://github.com/facebookresearch/esm | 20 |
| | ProteinMPNN | Message-passing neural network with three encoder and three decoder layers and 128 hidden dimensions that predicts protein sequences in an autoregressive manner from the N to C terminus using protein backbone features | Yes | No | Yes | ProteinMPNN achieved a sequence recovery rate of 52.4% when designing sequences for native protein backbones, outperforming Rosetta, which had a recovery rate of 32.9% | Yes | https://github.com/dauparas/ProteinMPNN. | 19 |
| | IgDesign | Model similar to AbMPN | Yes (The model was fine-tuned on specific antibody-antigen complexes, utilizing IgMPNN | Yes | Yes (antigen sequence and antibody framework (FWR) sequences are | For HCDR123 design, the model outperforms this HCDR3-only baseline on 7 out of 8 antigens. | Yes (Designed antibodies were tested in vitro against 8 therapeutic antigens confirming | https://github.com/AbSciBio/igdesign | 125 |
| | | | (both pretrained and fine-tuned) as its structural encoder to enhance the encoding of these complex interactions.) | | provided as context during training) | IgDesign outperforms IgMPNN on LCDR1 (100-shot amino acid recovery) and LCDR3 (1-shot amino acid recovery and 100-shot amino acid recovery). For 5/8 antigens, binders matched or exceeded reference antibody affinities. | binding rates up to 96.3% for HCDR3 designs. Binding affinities were experimentally determined, with 5 of 8 antigens showing affinities comparable to or better than reference antibodies.) | | |
| | AbMPNN | Based on ProteinMPNN framework, a structured transformer that utilizes a message-passing neural network (MPNN) for encoding structural features of proteins, optimized here specifically for antibodies. | Yes (trained on SAbDab dataset and Immunobuilder dataset) | Yes (fine-tuned on SAbDab and the OAS) | Yes (trained on full database for antibodies in complex with a protein antigens) | The model, compared to the ProteinMPNN model, improved designability by reducing RMSD for the CDR-H3 loop by 20% and increased sequence recovery rates across CDR loops from ~40% to ~60%. It also doubled the stability of generated antibodies, with 40% within 5 kcal/mol of native interface energy, and ensured all sequences were valid antibody structures, whereas 16.8% were invalid in ProteinMPNN | No | https://zenodo.org/records/8164693 can be run using code published on https://github.com/dauparas/ProteinMPNN | 94 |
| | AntiFold | Larger pre-trained ESM-IF1 architecture (142M parameters) | Yes (trained on SAbDab dataset and Immunobuilder dataset)) | Yes (fine-tuned on SAbDab and the OAS) | Yes (trained on full database for antibodies in complex with a protein antigens) | Achieved 75–84% sequence recovery across CDR regions, outperforming AbMPNN (63–76% excluding CDRH3). For framework regions, it recovered sequences with 87–94% accuracy, slightly better than AbMPNN (85–89%). Predicted sequences maintained structural similarity to experimental counterparts with a median RMSD of 0.67 Å for CDR regions, compared to 0.74 Å for AbMPNN, 0.75 Å for ESM-IF1 and 0.48 Å for | No | https://opig.stats.ox.ac.uk/data/downloads/AntiFold/ | 95 |

| | | | | | | native RMSD. The model achieved lower perplexity (better amino acid prediction) in the CDRH3 loop, averaging 2–8 mutations likely to preserve the fold versus 3–10 for AbMPNN. | | | |
|---|---|---|---|---|---|---|---|---|---|
| Developability | Ammeur et al. | Modified Wasserstein GAN (WGAN) architecture with gradient penalty | Yes (Sequences were structurally aligned using the AHo numbering system, simplifying the capture of structural relationships. Additionally, surface properties such as negative patch sizes are modeled to control aspects that impact antibody behavior and stability) | Yes (trained on over 400,000 human antibody sequences) | No (generation based on desirable biophysical properties of antibodies | The GAN model generated 100,000 diverse antibody sequences with a KL divergence of 0.57, closely matching human repertoires. It also reduced immunogenicity by 76%. | Yes (A library of 100,000 GAN-generated antibodies was expressed and evaluated via phage display. Select antibodies from this library were further validated for stability and developability by expression in CHO cells and various biophysical assays) | Not mentioned | 101 |
| | p-IGgen | Auto-regressive decoder-only language model using a GPT-2-like architecture | Yes (all 1.8M paired sequences were structurally modeled using ABB2 and then ran through TAP | Yes (trained on 1.8M paired VH/VL sequences taken from OAS) by finetuning) | No | Model achieved a Pearson correlation of 0.53 for immunogenicity prediction, a VH/VL mutation correlation of 0.52 (natural: 0.51), and generated sequences with high diversity and a mean Hamming distance of 11.3, demonstrating its ability to produce realistic, biologically plausible antibodies | No | github.com/oxpig/p-IgGen | 102 |
| | Hutchins et al. DeepAb | Model based on DeepAb. This prediction framework integrates Rosetta minimization, enhancing the spatial configuration of antibodies through computational design | Yes | Yes (it employs metrics tailored to antibody characteristics, including a specialized DeepAb score | No (it predicts mutation impacts on binding affinity based on an internal scoring (DeepAb) | 91% of variants showed improved thermal and colloidal stability, 94% exhibited improved affinity. The model showed 6%–35% success rates for different mutation combinations | Yes | https://github.com/RosettaCommons/DeepAb | 103 |

| | | | | | | that were projected based on computational scoring, compared to ~7% for random selection. | | | |
|---|---|---|---|---|---|---|---|---|---|
| | One-shot developable antibody design | The inverse folding model AbMPNN was utilized to generate sequences that maintain the structural integrity of antibodies, ensuring compatibility with target structures. Additionally, ESM guided specific mutations to enhance both binding affinity and developability. | Yes | Yes (Model focuses on enhancing key antibody properties like binding affinity and stability, tailoring its approach to meet antibody design requirements.) | Yes (Model began with established binders) | The pipeline achieved a 79% hit rate (57/72 designs) for binding and developability improvements, 48% (31/65) for escape mutation rescue, and generated binders with up to 74 sequence edits, showcasing efficiency and precision. | Yes | https://github.com/Exscientia/ab-characterisation; https://doi.org/10.5281/zenodo.13862717 | 104 |